\documentclass[12pt]{article}

\usepackage{cite}

\usepackage{times}

\usepackage{amsmath}
\usepackage[linesnumbered, ruled, vlined]{algorithm2e}
\usepackage{multicol}
\usepackage{multirow}

\title{GPU Accelerated AES Algorithm}

\author{
	{\footnotesize Canhui Wang $^{\dagger}$, Xiaowen Chu $^{\dagger,\star}$} \\
	{\footnotesize \textit{$^{\dagger}$ Computer Science, Hong Kong Baptist University, Hongkong, China}}\\
	{\footnotesize \textit{$^{\star}$ HKBU Institute of Research and Continuing Education, Shenzhen, China}}\\
	{\footnotesize \textit{\{chwang, chxw\}@comp.hkbu.edu.hk}}
}

\date{}

\begin{document} 

\baselineskip13pt
\maketitle

\begin{abstract}
	
	It has been widely accepted that Graphics Processing Units (GPUs) is one of promising schemes for encryption acceleration, in particular, the support of complex mathematical calculations such as integer and logical operations makes the implementation easier; however, complexes such as parallel granularity, memory allocation still imposes a burden on real world implementations.
	
	In this paper, we propose a benchmarking approach for AES accelerations, including both encryption and decryption. Specifically, we adapt the Electronic Code Book (ECB) mode for cryptographic transformation, T-boxes scheme for fast lookups, and a granularity of '\textit{one state per thread}' for thread scheduling. Our benchmarking results offer researchers a good understanding on GPU architectures and software accelerations. In addition, both our source code and experimental results are freely available.

\end{abstract}

\section{Introduction}

A graphics processing unit (GPU) is a specialized electronic circuit, which is originally designed to rapidly manipulate, and alter memory to accelerate the creation of images in a frame buffer. Nowadays GPUs are very efficient at manipulating computer graphics and image processing as well. Also their highly parallel structure makes them more efficient than general CPUs in the related fields such as statistical mechanics, mathematical biology and information security where the processing of large blocks of data is done in parallel. Compute Unified Device Architecture (CUDA), a parallel computing platform and application programming interface model developed by NVIDIA, is the earliest widely adopted programming model for GPU computing. To make it more easier for specialists in parallel programming, the CUDA platform is designed to work with programming languages such as C, C++, Fortran, MATLAB and etc. Additionally, CUDA supports many programming frameworks like OPENACC, OpenCL and etc. It has become one of the most popular choices for the implementation of computationally demanding algorithms.


\section{Related Work}

Abdelrahman et al., \cite{abdelrahman2017high} propose an implementation of the AES-128 ECB Encryption on three different GPU architectures (Kepler, Maxwell and Pascal). The results show that encryption speeds with
207 Gbps on the NVIDIA GTX TITAN X (Maxwell) and 280 Gbps on the NVIDIA GTX 1080 (Pascal) have been achieved by performing new optimization techniques using 32bytes/thread granularity.

Chengjian et al., \cite{liu2018g} propose a graphics processing unit (GPU)-based implementation of erasure coding named G-CRS which employs the Cauchy Reed-Solomon (CRS) code to overcome the aforementioned bottleneck. To maximize the coding performance of G-CRS, they designed and implemented a set of optimization strategies, such as a compact structure to store the bitmatrix in GPU constant memory, efficient data access through shared memory, and decoding parallelism, to fully utilize the GPU resources.

Kaiyong et al., \cite{zhao2014g} develope G-BLASTN, a GPU-accelerated nucleotide alignment tool based on the widely used NCBI-BLAST. G-BLASTN can produce exactly the same results as NCBI-BLAST, and it has very similar user commands. Compared with the sequential NCBI-BLAST, G-BLASTN can achieve an overall speedup of 14.80X under ‘megablast’ mode. And they \cite{zhao2010gpump} propose  to exploit the computing power of Graphic Processing Units (GPUs) for homomorphic hashing. Specifically, they demonstrate how to use NVIDIA GPUs and the Computer Unified Device Architecture (CUDA) programming model to achieve 38 times of speedup over the CPU counterpart. They also develop a multi-precision modular arithmetic library on CUDA platform, which is not only key to our specific application, but also very useful for a large number of cryptographic applications. Also, they \cite{li2012implementation} propose a GPU based AES implementation of which the frequently accessed T-boxes were allocated on on-chip shared memory and the granularity that one thread handles a 16 bytes AES block was adopted. They achieved the performance of around 60 Gbps throughput on NVIDIA Tesla C2050 GPU, which runs up to 50 times faster than a sequential implementation based on Intel Core i7-920 2.66GHz CPU.

Iwai et al., \cite{iwai2012acceleration} present results of several experiments that were conducted to elucidate the relation between memory allocation styles of variables of AES and granularity as the parallelism exploited
from AES encoding processes using CUDA with an NVIDIA GeForce GTX285 (Nvidia Corp.).
Results of these experiments showed that the 16 bytes/thread granularity had the highest performance.
It achieved approximately 35 Gbps throughput.

Xinxin et al., \cite{mei2014benchmarking, mei2017dissecting} propose a novel fine-grained benchmarking approach and apply it on two popular GPUs, namely Fermi and Kepler, to expose the previously unknown characteristics of their memory hierarchies. They also investigate the impact of bank conflict on shared memory ac- cess latency.

Xiaowen et al., \cite{chu2009practical, chu2008practical} exploit the potential of the huge computing power of Graphic Processing Units (GPUs) to reduce the computational cost of network coding and homomorphic hashing. With their network coding and HHF implementation on GPU, they observed significant computational speedup in comparison with the best CPU implemen- tation. Their implementation can lead to a practical solution for defending against the pollution attacks in distributed systems.

Ahmed et al., \cite{abdelrahman2017analysis} propose an AES-128 algorithm (ECB mode) implementation on three different GPU architectures with different values of granularities (32,64 and 128 bytes/thread). Their results show that the throughput factor reaches 277 Gbps, 201 Gbps and 78 Gbps using the NVIDIA GTX 1080 (Pascal),
the NVIDIA GTX TITAN X (Maxwell) and the GTX 780 (Kepler) GPU architectures.

Conti et al., \cite{conti2017design} presents a direct comparison between FPGA and GPU used as accelerators for the AES cipher. The results achieved on both platforms and their analysis has been compared to several others in order to establish which device is best at playing the role of hardware accelerator by each solution showing interesting considerations in terms of throughput, speedup factor, and resource usage. Their analysis suggests that, while hardware design on FPGA remains the natural choice for consumer-product design, GPUs are nowadays the preferable choice for PC based accelerators, especially when the processing routines are highly parallelizable.

Ma et al., \cite{ma2017implementation} implement AES decryption in CBC mode, a mode that is widely used by many applications, on GPU using Cuda, a framework developed by NVIDIA and friendly to use. To achieve the best performance, they give a comprehensive performance analysis to the implementations based on GPU under different parameters setting that include the size of input data, the number of threads per block, memory allocation style and parallel granularity. Through experiment evaluation based on our implementation, the best performance of AES on GPU is 112 times over the serial AES algorithm on CPU. 


\section{Preliminary}

\subsection{Block Cipher Mode}

Block cipher mode is a schema that uses a block cipher to encrypt messages of arbitrary length in a way that provides confidentiality and authenticity. Many block cipher modes \cite{black2002block, dworkin2016recommendation} have been well defined. Five of them will be discussed below.

\subsubsection{The Electronic Codebook Mode}

The Electron Code book (ECB) is a mode that assigns a fixed cipher text block to each plaintext block, which can be defined as the equation (1).

\begin{equation}
C_{i} = ciphers\left ( P_i \right )
\end{equation}

\noindent where $i = 1, 2, 3, ..., n$. The plain text $P$ can be represented as $P = P_{1} P_{2} P_{3}...P_{n}$, and the cipher text $C$ can be represented as $C=C_{1} C_{2} C_{3}...C_{n}$, respectively.

\subsubsection{The Cipher Block Chaining Mode}

The Cipher Block Chaining (CBC) is a mode where the previous cipher-text blocks are combined with the plain-text blocks. An initialization vector is required to combine with the first plain-text block. The encryption in CBC mode can be defined as the equation (2).

\begin{equation}
	C_{i} = P_i \oplus C_{i-1}
\end{equation}

\noindent where $i=1,2,3,...,n.$ The plain text $P$ can be represented as $P=P_{1}P_{2}P_{3}...P_{n},$ and the cipher text $C$ can be represented as $C=C_{1}C_{2}C_{3}...C_{n}$. In particular, the initialization vector $C_0$ is not necessary to be secret but it should be unpredictable.

\subsubsection{The Cipher Feedback Mode}

The Cipher Feedback (CFB) is a mode that is similar to the previous described CBC mode. A main difference is that the CFB mode is fed with the cipher-text data (not the plain-text data) from previous rounds. The encryption in CFB mode can be defined as the equation (3).

\begin{equation}
	C_{i} = P_{i} \oplus \rm{CFB}_{k} \left ( C_{i-1} \oplus P_{i-1} \right)
\end{equation}   

\noindent where $i=1,2,3,...,n.$ The plain text $P$ can be represented as $P=P_{1}P_{2}P_{3}...P_{n},$ and the cipher text $C$ can be represented as $C=C_{1}C_{2}C_{3}...C_{n}$. If any one single bit of a plain text $P_i$ is damaged, the output cipher block will be damaged.

\subsubsection{The Output Feedback Mode}

The Output Feedback (OFB) is a mode where key-stream bits $\rm{IV}$ are created for data encryption. The encryption of OFB mode can be defined as the equation (4).

\begin{equation}
	C_{i} = P_{i}  \oplus \rm{OFB} \left ( \rm{IV}_{i-1} \right)
\end{equation}
 
\noindent where $i=1,2,3,...,n.$ The plain text $P$ can be represented as $P=P_{1}P_{2}P_{3}...P_{n},$ and the cipher text $C$ can be represented as $C=C_{1}C_{2}C_{3}...C_{n}$.

\subsubsection{The Counter Mode}

The Counter (CTR) mode makes block cipher the way similar to OFB mode. The key-stream bits are created regardless of the content of encrypted blocks data. In such a mode, subsequent values of an increasing counter are added to a unique number and then the plain text data are encrypted as usual. The encryption of CTR mode can be defined as the equation (5).

\begin{equation}
	C_{i} = P_{i} \oplus CTR \left (R+i\right)
\end{equation} 

\noindent where $R=\left \{ 0, 1, 2,...,2^{\left | P \right |}-1 \right  \}$, $i=1,2,3,...,n.$ The plain text $P$ can be represented as $P=P_{1}P_{2}P_{3}...P_{n},$ and the cipher text $C$ can be represented as $C=C_{1}C_{2}C_{3}...C_{n}$.

\subsubsection{Comparison}


\begin{table}[htbp]
	\centering
	\caption{Comparison }
	{\footnotesize \begin{tabular}{l|l}
		\hline
		Block Cipher Mode & Parallel Capacity \\ \hline
		ECB               & Suitable          \\ \hline
		CBC               & Unsuitable        \\ \hline
		CFB               & Unsuitable        \\ \hline
		OFB               & Unsuitable        \\ \hline
		CTR               & Suitable          \\ \hline
	\end{tabular}}
\end{table}    

Table 1 shows the parallel capacity of five block cipher modes. As we can see, both ECB and CTR modes naturally support parallel computing because every one single block of them can be encrypted independently. In the following sections, we will focus on ECB mode and analyze the implementation of ECB based AES algorithms.

\subsection{AES Basics}

It is widely acknowledged that Rijndael is a government encryption standard that has been submitted to the International Organization for Standardization (ISO) and the Internet Engineering Task Force (IETF) as well as the Institute of Electrical and Electronics Engineers (IEEE) as an approval encryption standard. One of the main factors for its succeed acceptance is that it supports parallel implementations on a wide range of platforms. In this section, we will discuss both the theory and the construction of AES.

\subsubsection{The Field of $\rm{GF}\left ( 2^{n} \right )$}

$\rm{GF}\left ( 2^{n} \right )$ is a finite field that contains $2^{n}$ elements. It supports some basic mathematical operations like multiplication, addition and etc. Elements in a finite field can be represented in several different ways such as polynomial expressions, integer numbers etc. In AES, the field of $\rm{GF}\left ( 2^{8} \right )$ is adopt. To describe a finite field in detail, a simple example of $\rm{GF}\left ( 2^{3} \right )$ is given in Table 2.

\begin{table}[htbp]
	\centering
	\caption{All elements in the finite field of $\rm{GF}\left ( 2^{3} \right )$}
	{\footnotesize \label{my-label}
	\begin{tabular}{l|l|l}
		\hline
		Polynomial & Tri-Tuple & Integer \\ \hline
		0          & $\left ( 0,0,0 \right )$        & 0       \\ \hline
		1          & $\left ( 0,0,1 \right ) $        & 1       \\ \hline
		$x$          & $\left ( 0,1,0 \right )$         & 2       \\ \hline
		$x+1$        &$ \left ( 0,1,1 \right )$         & 3       \\ \hline
		$x^2$         & $\left ( 1,0,0 \right ) $        & 4       \\ \hline
		$x^{2}+1$       & $\left ( 1,0,1 \right )$         & 5       \\ \hline
		$x^{2}+x$       & $\left ( 1,1,0 \right )  $       & 6       \\ \hline
		$x^{2}+x+1$     & $\left ( 1,1,1 \right ) $        & 7       \\ \hline
	\end{tabular}}
\end{table}

Table 2 shows all elements in the field of $\rm{GF}\left ( 2^{3} \right )$ in three different ways: polynomial expressions, tri-tuple and integer numbers. In general, all elements of a finite filed $\rm{GF}\left ( 2^{n} \right )$ can be represented in equation (6).

\begin{equation}
	\rm{GF}\left ( 2^{n} \right ) = a_{i}x^i
\end{equation}

\noindent where $i=0,1,2,3,...,n-1$, $a_{i}=0$ or $1$. In particular for AES, $n=8$.

\noindent \textbf{Definition 1. Addition:} in the polynomial representation, the addition of two elements in a finite field simply means adding the coefficients of the two elements. For example, given $f(x)=x^2+x$, $g(x)=x^2+x+1$, then $f(x)+g(x)$ can be represented in equation (7).

\begin{equation}
	(x^2+x) +(x^2+x+1) =1
\end{equation}

\noindent \textbf{Definition 2. Multiplication:} in the polynomial representation, the multiplication of two elements in a finite field is similar to normal polynomial multiplication, the only difference is that an irreducible polynomial is required for finite filed multiplication. An irreducible polynomial is a non-constant polynomial that cannot be factored into the product of two non-constant polynomials. Specifically, in AES the irreducible polynomial denoted as $m(x)$ as given in equation (8).

\begin{equation}
	m(x) = x^8+x^4+x^3+x+1
\end{equation}

For example, given $f(x)=x^6+x^4+x^2+x+1$,$g(x)=x^7+x+1$, then $f(x)g(x)$ can be reasoned as follows,

\begin{equation}
	(x^6+x^4+x^2+x+1)(x^7+x+1)
\end{equation}

\begin{equation}
=x^{13}+x^{11}+x^9+x^8+x^7+x^7+x^5+x^3+x^2+x^6+x^4+x^2+x^1+x^0
\end{equation}

\begin{equation}
=x^{13}+x^{11}+x^9+x^8+x^6+x^5+x^4+x^3+x^0
\end{equation}

\noindent and

\begin{equation}
(x^{13}+x^{11}+x^9+x^8+x^6+x^5+x^4+x^3+x^0)mod (m(x))
\end{equation}

\begin{equation}
=x^7+x^6+1
\end{equation}

\noindent thus

\begin{equation}
f(x)g(x)=x^7+x^6+1
\end{equation}

\noindent \textbf{Definition 3. xtime():} in AES, the multiplication by $02$ in decimal (or $x$ in polynomial) is denoted as $xtime()$. For example, if we multiply $f(x)$ by the polynomial $x$, we can reason it as follows,

\begin{equation}
xtime(f(x))=xf(x)
\end{equation}

\begin{equation}
=x^7+x^5+x^3+x^2+x^1
\end{equation}

\noindent thus,

\begin{equation}
xtime(f(x))=x^7+x^5+x^3+x^2+x^1
\end{equation}

One implementation of $xtime()$ is to take a fixed number of cycles which are independent from the value of its arguments; however, such an approach suffers the power analysis attacks \cite{fontana2014method, alioto2014effectiveness, luo2015towards}. A better approach is to define look-up tables denoted by $M$, where $M[a]=xtime(a)$. All elements of AES can be written as a sum of powers of $x$, thus it is clear that multiplication by any value in AES could be implemented by a repeated use of look-up table $M$. Note that the function of $xtime()$ equals to a shift operation and a conditional $XOR$ operation. Therefore, the function $xtime()$ is helpful to improve the performance of AES.

\subsubsection{AES Specification}

\noindent \textbf{Definition 4. sub-bytes():} in AES, the sub-bytes() transformation is a non-linear operation where each byte of a state is operated independently. In addition, to accelerate the speed of AES, a 256-byte loop-up table named sbox \cite{coron2016faster} is used during the process of encryption and decryption. The sub-bytes() transformation can be represented as follows.

\begin{equation}
	state_{i,j} = sbox_t 
\end{equation}

\noindent where $t=state_{i,j}$, $0 \leq i,j \leq 3 $. The equation (18) shows the sub-bytes() operation where the state matrix is represented as $state$ and the sbox is represented as $sbox$. Obviously, the sub-bytes() operation can be implemented in parallel by operating on each byte of the state matrix.

\noindent \textbf{Definition 5. shift-rows():} in AES, the shift-rows() operation is to shift rows of a state. In general, the rows of the state matrix are cyclically shifted over different offsets. In this case, the original state matrix is denoted as $state$, the state matrix after shifting rows is denoted as $state'$. Then, the shift-rows() operation can be expressed in equation (19).

\begin{equation}
	state _{i,j}^{'} = state_{i,t}
\end{equation}

\noindent where $t=(i+j)mod(4)$. After the shift-rows() operation, all four bytes of each specific column are spread over four distinct columns. Note that the shift-rows() operation heavily depends on the round key.

\noindent \textbf{Definition 6. mix-columns():} assume that the state matrix $state$, after the operation of shifting rows, the state matrix is denoted by $state'$. In addition, a fixed matrix is denoted by $A$. Then, the mix-columns operation can be expressed in equation (20).

\begin{equation}
state' = A \otimes state \\ 
\end{equation}

$$
	A=\begin{bmatrix}
	0x02  & 0x03 & 0x01 & 0x01 \\ 
	0x01  & 0x02 & 0x03 & 0x01\\ 
	0x01  & 0x01 & 0x02 & 0x03\\ 
	0x03  & 0x01 & 0x01 & 0x02
	\end{bmatrix}
$$	

\noindent where $A$ is a fixed matrix representing the rows shifting operation.

\noindent \textbf{Definition 7. add-roundkey():} a round key is applied to the state by a simple bitwise operation. Let assume that the state matrix $state$, after the operation of adding round key, the state matrix is denoted as $state'$ and the expanded round key used in this step is denoted as $epdkey$. Then the add-roundkey() operation can be expressed in equation (21).

\begin{equation}
	state' = state \oplus  epdkey
\end{equation}

Note that one of the most important things is about how to generate and use round keys. The round keys are generated from the input cipher key by Key Schedule \cite{huang2016transposition}. The key schedule involves two steps: the key expansion and round key selection. In the first step, the total length of twelve round keys is equal to the block length that multiplies the number of rounds. For example, for a block with 128 bits block size and 10 rounds, the total length round keys (or expanded key) is $128*(10+1)=1408$ bits. In the second step, round keys are sequentially taken from the expanded key. For example, the first round key consists of the first $Nb$ words in the expanded key, the second round key consists of the second $Nb$ words in the expanded key, so on and so forth.

\subsubsection{T Tables}

So far, all specific operations of AES, including sub-bytes(), shift-rows(), mix-columns(), add-roundkey(), have been discussed (also see Algorithm 1).

\begin{algorithm}[htbp]
	{\footnotesize \SetKwInOut{Input}{Input}\SetKwInOut{Output}{Output}
		\Input{a state, a cipherkey.}
		\Output{an encrypted state.}
		\BlankLine
		
		keyExpansion(cipherkey, expandedkey); \\
		AddRoundKey(state, expandedkey[0]); 
					

		\For{int i = 0$;$ i$<$Nr$;$ i++}{
			sub-bytes(state); (see \textbf{\textit{Definition 4}})
			
			shift-rows(state);(see \textbf{\textit{Definition 5}})
			
			mix-columns(state); (see \textbf{\textit{Definition 6}})
			
			add-roundkey(state, expandedkey[i]); (see \textbf{\textit{Definition 7}})
		}
					
		sub-bytes(state); (see \textbf{\textit{Definition 4}})\\
		shift-rows(state); (see \textbf{\textit{Definition 5}})\\
		mix-columns(state); (see \textbf{\textit{Definition 6}})\\
		add-roundkey(state, expandedkey[Nr]); (see \textbf{\textit{Definition 7}})			
		
%
%
%
%

}
	
	\caption{{\small An overview of AES implementation}}
\end{algorithm}

\begin{equation}
	T_{0}(w)
	\begin{bmatrix}
	xtime(sbox(w), 02)\\ 
	sbox(w)\\ 
	sbox(w)\\ 
	xtime(sbox(w), 03)
	\end{bmatrix}
\end{equation}

\begin{equation}
	T_{1}(w)
	\begin{bmatrix}
	xtime(sbox(w), 03)\\ 
	xtime(sbox(w), 02)\\ 
	sbox(w)\\ 
	sbox(w)
	\end{bmatrix}
\end{equation}

\begin{equation}
	T_{2}(w)
	\begin{bmatrix}
	sbox(w)\\ 
	xtime(sbox(w), 03)\\ 
	xtime(sbox(w), 02)\\ 
	sbox(w)
	\end{bmatrix}
\end{equation}

\begin{equation}
	T_{3}(w)
	\begin{bmatrix}
	sbox(w)\\ 
	sbox(w)\\ 
	xtime(sbox(w), 03)\\ 
	xtime(sbox(w), 02)
	\end{bmatrix}
\end{equation}

However, in order to achieve a higher performance, four look-up tables named T Tables are introduced. Note that for each AES round, T boxes consists of round transformations including sub-bytes, shift-rows, mix-columns and add-roundkey. In other words, a T Table represents a round transformation. All four T Tables are computed as equations (22, 23, 24 and 25), where $w=0,1,2,3,...,255.$ Consider a given round input $p$, the round output could be represented as follows,

\begin{equation}
	e_j=T_0[p_{0,j}] \oplus T_1[p_{1,j+1}] \oplus T_2[p_{2,j+2}] \oplus T_3[p_{3,j+3}] \oplus k_j
\end{equation}

\noindent where $k_j$ represents the $j-th$ column of a round key.

\section{GPU Accelerated AES}

\subsection{Parallel Granularity}
Our GPU Accelerated AES adapts the 'one state per thread' parallel granularity which means that each thread is mapped to a AES state. Messages are divided into multiple 16-byte states and these states are executed by GPUs synchronously. Note that each GPUs thread in this case can be completed independently since states of different threads do not have interdependencies, meaning that any thread do not need to wait for other threads.

\subsection{T-Tables Allocation}

T Tables are essentially look-up tables through which users can quickly implement part of cryptographic operations. Every thread needs to get access T tables in each round of cryptographic operations. Hence we need to load T tables in the memory of GPU in advance. In our benchmarking approach, we load T tables into CUDA constant memory. Another possible solution is to load T tables into CUDA share memory or registers.

\subsection{Round-keys Allocation}

AES mainly guarantees the security of the password through the confidentiality of the key and the length of the key. In the implementation of  a specific AES algorithm, each round of AES encryption requires key operations. Each thread of CUDA needs an input of a different round key. Of course, tens of different round keys are generated from one single input key. In this case, we put round keys in the constant memory. All threads share user round keys in the constant memory.

\subsection{Plaintext Allocation}

In the AES of ECB mode, the plaintext information is divided into several different states/blocks to be encrypted. In our benchmarking approach, we use 'one thread per state' method. For example, if we have 100 states, that is 1600 bytes, then we need at least 100 GPU threads. After each thread acquires 16 bytes of data, the plaintext data is stored in the thread's register. Register data between threads is inaccessible to each other. After all threads complete the data encryption, we get back the encrypted data from GPU memory.

\section{Experimental Results}

Table 3 shows the configuration of our experiment platform. We conduct our experiments on a Ubuntu 16.04 LTS 64-bit operating system with a 12-GiB memory, an Intel Core i5-7200U @ 2.5GHz*4, an GeForce 940MX/PCIe/SSE2 GPU. The version of nvcc compliation is V7.5.17 and the version of GCC is V5.4.0. Our CPU code is written in standard C language and the GPU code is written in CUDA.

\begin{table}[htbp]
	\centering
	\caption{Environment specification}
	{\footnotesize \label{my-label}
	\begin{tabular}{l|c}
		\hline
		CPU              & Intel Core i5-7200U @ 2.5GHz*4 \\ \hline
		GPU              & GeForce 940MX/PCIe/SSE2        \\ \hline
		Memory           & 12 GiB                         \\ \hline
		OS               & Ubuntu 16.04 LTS, 64 bits      \\ \hline
		CUDA compliation & V7.5.17                        \\ \hline
		GCC              & V5.4.0                         \\ \hline
	\end{tabular}}
\end{table}

Table 4 shows the performance between CPU based AES encryption and GPU based AES encryption. When the file size is small, say 1024 bytes, CPU performs better than GPU; however, when the file size is getting larger, say 10 KBytes, GPU performs much better than CPU. The performance of CPU is relatively stable while the performance of GPU can be highly excavated, meaning that GPU performance can be better used when more concurrent tasks are available. In particular, when the file size is 1190402 bytes, the CPU's encryption performance is still maintained at about 7510103 bytes per second while the GPU's performance has increased to 223928517 bytes per second.

\begin{table}[htbp]
	\centering
	{\footnotesize 	\caption{AES encryption CPU v.s. GPU}
		\label{my-label}
		\begin{tabular}{l|l|l|l|l}
			\hline
			\multirow{3}{*}{File Size (bytes)} & \multicolumn{4}{c}{Encryption}                                                                          \\ \cline{2-5} 
			& \multicolumn{2}{c|}{Time (seconds)}                 & \multicolumn{2}{c}{Throughput (bytes per second)} \\ \cline{2-5} 
			& \multicolumn{1}{c|}{CPU} & \multicolumn{1}{c|}{GPU} & \multicolumn{1}{c|}{CPU} & \multicolumn{1}{c}{GPU} \\ \hline
			1202                               & 0.000152                 & 0.000732                 & 7921052.63               & 1644808.74              \\ \hline
			4652                               & 0.000689                 & 0.000770                 & 6769230.77               & 6057142.85              \\ \hline
			9302                               & 0.001418                 & 0.000784                 & 6564174.89               & 11872448.97             \\ \hline
			18602                              & 0.002545                 & 0.000807                 & 7313163.06               & 23063197.03             \\ \hline
			37202                              & 0.004985                 & 0.000886                 & 7463189.57               & 41990970.65             \\ \hline
			74402                              & 0.010099                 & 0.001048                 & 7367462.12               & 70996183.21             \\ \hline
			148802                             & 0.020109                 & 0.001302                 & 7399870.70               & 114288786.48            \\ \hline
			297602                             & 0.039651                 & 0.001896                 & 7505586.24               & 156964135.02            \\ \hline
			595202                             & 0.079503                 & 0.003145                 & 7486560.26               & 189254054.05            \\ \hline
			1190402                            & 0.158507                 & 0.005316                 & 7510103.65               & 223928517.68            \\ \hline
	\end{tabular}}
\end{table}

Table 5 shows the performance between CPU based AES decryption and GPU based AES decryption. When the file size is small, say 1024 bytes, CPU performs better than GPU; however, when the file size is getting larger, say 10 KBytes, GPU performs much better than CPU. In particular, when the file size is 1190402 bytes, the CPU's decryption performance is still maintained at 5239385 bytes per second while the GPU's performance has increased to 203592269 bytes per second.

\begin{table}[htbp]
	\centering
	\caption{AES decryption CPU v.s. GPU}
	{\footnotesize 
		\label{my-label}
		\begin{tabular}{l|l|l|l|l}
			\hline
			\multirow{3}{*}{File Size (bytes)} & \multicolumn{4}{c}{Decryption}                                                                          \\ \cline{2-5} 
			& \multicolumn{2}{c|}{Time (seconds)}                 & \multicolumn{2}{c}{Throughput (bytes per second)} \\ \cline{2-5} 
			& \multicolumn{1}{c|}{CPU} & \multicolumn{1}{c|}{GPU} & \multicolumn{1}{c|}{CPU} & \multicolumn{1}{c}{GPU} \\ \hline
			1202                               & 0.000264                 & 0.000771                 & 4560606.06               & 1561608.30              \\ \hline
			4652                               & 0.001168                 & 0.000821                 & 3993150.68               & 5680876.98              \\ \hline
			9302                               & 0.002081                 & 0.000790                 & 4472849.59               & 11782278.48             \\ \hline
			18602                              & 0.003889                 & 0.000821                 & 4785806.12               & 22669914.74             \\ \hline
			37202                              & 0.007196                 & 0.000918                 & 5170094.49               & 40527233.12             \\ \hline
			74402                              & 0.014296                 & 0.001040                 & 5204532.74               & 71542307.69             \\ \hline
			148802                             & 0.028493                 & 0.001331                 & 5222475.69               & 111798647.63            \\ \hline
			297602                             & 0.056839                 & 0.002018                 & 5235911.96               & 147474727.45            \\ \hline
			595202                             & 0.113610                 & 0.003577                 & 5239010.65               & 166397539.84            \\ \hline
			1190402                            & 0.227203                 & 0.005847                 & 5239385.04               & 203592269.54            \\ \hline
	\end{tabular}}
\end{table}

\section{Conclusion}

In this paper, we propose a benchmarking approach for GPU based AES implementation involving both encryption and decryption progress. We adapt the Electronic Code Book (ECB) mode for cryptographic transformation, T-boxes scheme for fast lookups, and a granularity of '\textit{one state per thread}' for thread scheduling. Our benchmarking results offer researchers a good understanding on GPU architectures and software accelerations. Our experimental results show that GPU has significantly advantages than CPU in performance when using AES for encrypting (or decrypting) large files, say more than 4K bytes. And when the size of parallelizable files is getting larger, the performance of GPU can be fully utilized. In addition, both our source code and experimental results are freely available at github (https://github.com/Canhui/AES-ON-GPU).

\section*{Acknowledgements}
This work is supported by Shenzhen Basic Research Grant SCI-2015-SZTIC-002.

{\small \bibliography{scibib}}

\bibliographystyle{ieeetran}

\end{document}